\documentstyle[12pt]{article}

\begin{document}
\newcommand\til[1] {\buildrel \sim \over {#1}}
\newcommand\fr[2] {\mbox{$\,{#1\over #2}\,$}}
\newcommand\FR[2] {\;{\strut\displaystyle{#1} \over
 \strut\displaystyle{#2}}\;}
\newcommand\tr {{\rm~tr}}
\renewcommand\thefootnote {\fnsymbol{footnote})~}

\setcounter{page}0
\thispagestyle{empty}
{}~\vfill
\begin{center}
 {\Large\bf Antisymmetric tensor matter fields:\\
  an abelian model}\\[1cm]
 {\large L.V.Avdeev}\\[5mm]
 {\em Department of Physics, The University of Pisa,\\
  2 Piazza Torricelli, I-56100 Pisa, Italy\\ and\\
  Laboratory of Theoretical Physics, Joint Institute
  for Nuclear Research,\\
  141980 Dubna (Moscow Region), Russian Federation;
  \footnote{Permanent address}\\
  Electronic mail:}
 avdeevL@theor.jinrc.dubna.su\\[5mm]
 {\large and\\[5mm] M.V.Chizhov}\\[5mm]
 {\em Center of Space Research and Technologies,\\
  Faculty of Physics, University of Sofia,\\
  1126 Sofia, Bulgaria;\\
  Electronic mail:}
 physfac2@bgearn.bitnet
 \end{center} \vfill

\begin{abstract}
We present a simple renormalizable abelian gauge model
which includes antisymmetric second-rank tensor fields as
matter fields rather than gauge fields known for a long
time. The free action is conformally rather than gauge
invariant. The quantization of the free fields is
analyzed and the one-loop renormalization-group functions
are evaluated. Transverse free waves are found to convey
no energy. The coupling constant of the axial-vector
abelian gauge interaction exhibits asymptotically free
ultraviolet behavior, while the self-couplings of the
tensor fields do not asymptotically diminish.
\end{abstract}
\vfill
\pagebreak

\section{The lagrangian} \indent

The structure of the kinetic term and interactions for
antisymmetric tensor matter fields can be obtained by
dynamical generation \cite{Dyn} from fundamental spinors
like in the Nambu--Jona--Lasinio model. We shall not stop
at the details of that and proceed to studying the
resulting lagrangian which could as well be guessed in
some other way.

It should be understood from the very beginning that the
fields under consideration are {\em not} the
antisymmetric tensor {\em gauge} fields introduced a long
time ago \cite{Notof} by analogy with the electromagnetic
vector field. We are going to deal with the {\em matter}
fields, the free action for which is {\em conformally}
rather than gauge invariant. Such fields appeared in
extended conformal supergravity theories \cite{Conf}, but
they have not specifically been studied there in detail.

We present a simple abelian gauge model that includes
this new type of matter fields, analyze their
quantization, and compute the one-loop renormalizations
in the model.

The lagrangian density in four dimensions is of the form
\begin{eqnarray} {\cal L} &=&
 \fr 1 2 \left( \partial_\lambda T_{\mu\nu} \right)^2
 - 2 \left( \partial_\mu T_{\mu\nu} \right)^2
 - \fr 1 4
 \left( \partial_\mu A_\nu - \partial_\nu A_\mu
 \right)^2
 + {\rm i} \overline{\psi} \gamma_\mu \partial_\mu \psi
 \nonumber\\
&& +~ h~ \overline{\psi} \gamma_5 \gamma_\mu A_\mu \psi
 + 4 h~ A_\mu
 \left( T_{\mu\nu} \partial_\lambda \til{T}_{\lambda\nu}
  - \til{T}_{\mu\nu} \partial_\lambda T_{\lambda\nu}
 \right) \nonumber\\
&& +~ 4 h^2
 \left[ \fr 1 2 \left( A_\lambda T_{\mu\nu} \right)^2
  - 2 \left( A_\mu T_{\mu\nu} \right)^2
 \right] \nonumber\\
&& +~ y~ \overline{\psi} \sigma_{\mu\nu} T_{\mu\nu} \psi
 + \fr 1 4 q
 \left[ \fr 1 2 \left( T_{\mu\nu} T_{\nu\mu} \right)^2
  - 2 T_{\mu\nu} T_{\nu\rho} T_{\rho\lambda}
  T_{\lambda\mu}
 \right], \label{L} \end{eqnarray}
where $A_\mu$ is a real axial-vector gauge field,
$T_{\mu\nu}$=$-$$T_{\nu\mu}$ is a real tensor matter
field, and $\psi$ is a Dirac spinor field.

We denote $\til{T}_{\mu\nu}$=$\fr{1}2
\epsilon_{\mu\nu\alpha\beta}~ T_{\alpha\beta}$~,
$\sigma_{\mu\nu}$=$\fr{1}2$i$ [\gamma_\mu,\gamma_\nu]$,
use the $(+---)$ Min\-kow\-ski metric, and always imply
covariant contractions of repeated Lorentz greek indices
without distinguishing their positions. The $\epsilon$
tensor and $\gamma_5$ satisfy
\begin{eqnarray*} &\epsilon_{\mu_1\ldots\mu_4}~
 \epsilon_{\nu_1\ldots\nu_4} = -~ g_{\mu_1}{}^{[\nu_1}
 \ldots g_{\mu_4}{}^{\nu_4]},~~~~~~~
 \til{\tilde{T}}_{\mu\nu} = - T_{\mu\nu}~,& \\
&\tr
 \left( \gamma_5 \gamma_{\mu_1} \ldots \gamma_{\mu_4}
 \right) = -4 {\rm i}~ \epsilon_{\mu_1\ldots\mu_4}~,~~~~~
 \gamma_5^2={\bf 1},~~~~~
 \fr 1 2 {\rm i}~ \epsilon_{\mu\nu\alpha\beta}~
 \sigma_{\alpha\beta} = \gamma_5~ \sigma_{\mu\nu}~.&
 \end{eqnarray*}
There are two useful identities which follow from
antisymmetry of $T_{\mu\nu}$ and the definition of the
$\epsilon$ tensor:
\begin{equation}
 \til{T}_{\mu\lambda} \til{T}_{\lambda\nu} =
 T_{\mu\lambda} T_{\lambda\nu}
 + \fr 1 2 g_{\mu\nu} T_{\alpha\beta}^2~,~~~~~~
 T_{\mu\lambda} \til{T}_{\lambda\nu} = \fr 1 4
 g_{\mu\nu} T_{\alpha\beta} \til{T}_{\beta\alpha}~.
 \label{id} \end{equation}

It can directly be checked that our lagrangian (\ref{L})
is invariant, up to a total derivative, under the
infinitesimal abelian gauge transformations
\begin{eqnarray*} &\delta A_\mu = \partial_\mu \omega,
 ~~~~~~ \delta \psi = - {\rm i} h \omega~ \gamma_5 \psi,
 ~~~~~~ \delta \overline{\psi} = - {\rm i} h \omega~
 \overline{\psi} \gamma_5~,& \\
&\delta T_{\mu\nu} = - 2 h \omega
 \til{T}_{\mu\nu}~,~~~~~~ \delta \til{T}_{\mu\nu} =
 2 h \omega T_{\mu\nu}~.& \end{eqnarray*}
The ratio of the charges of the spinor and tensor fields
is fixed by requiring the invariance of the Yukawa-like
$y$ interaction term.

An essential difference of the antisymmetric tensor
matter fields from their gauge counterparts of
ref.\cite{Notof} is the possibility of introducing a
renormalizable self-interaction in an abelian model. Its
form, which is presented by the $q$ term in eq.(\ref{L}),
is uniquely determined by the gauge invariance.

It is also important to point out that no explicit mass
terms can be introduced because neither $M^2
T_{\mu\nu}^2$~, $M^2 T_{\mu\nu} \til{T}_{\nu\mu}$~, $m
\overline{\psi} \psi$, nor $m \overline{\psi} \gamma_5
\psi$ would be gauge-inva\-ri\-ant. The fields may
acquire masses owing to a spontaneous symmetry breaking,
for example, after adding scalar fields in the model. We
do not tackle this problem for the time being, although
it is of importance for physical applications
\cite{New}.

\section{Quantization of free fields} \indent

The canonical hamiltonian ${\cal H}= {\buildrel. \over
\varphi}~ \partial{\cal L}/ \partial{\buildrel. \over
\varphi} -{\cal L}$ for the free part of the tensor-field
lagrangian (\ref{L}) can be reduced to the following
form:
\begin{eqnarray} {\cal H}_T =
 \left( \partial_0 T_{0j} \right)^2
 - \left( \vec{\partial}~ T_{0j} \right)^2
 + 2 \left( \partial_j T_{0j} \right)^2 \nonumber\\*
 + \left( \partial_0 T_j \right)^2
 - \left( \vec{\partial}~ T_j \right)^2
 + 2 \left( \partial_j T_j \right)^2, \label{H}
 \end{eqnarray}
where $\vec{\partial}$ and latin indices refer to space
dimensions with the Euclidean summation implied, the
mixed time-space tensor components $T_{0j}$ form a
3-dimensional vector, and the purely space components
$T_{jk}= \epsilon_{jkl}~ T_l$ are parameterized by an
axial 3-vector $T_j$.

The classical equations of motion
$$ \partial_\lambda^2 T_{\mu\nu} + 2 \partial_\lambda
 \left( \partial_\mu T_{\nu\lambda}
  - \partial_\nu T_{\mu\lambda}
 \right) = 0 $$
can explicitly be rewritten in components as
\begin{eqnarray*}
 \left[ \delta_{jk}
  \left( \partial_0^2 + \vec{\partial}^2 \right)
  - 2 \partial_j \partial_k
 \right] T_{0k} &=&
 - 2 \epsilon_{jkl}~ \partial_0 \partial_k T_l~ , \\
 \left[ \delta_{jk}
  \left( \partial_0^2 + \vec{\partial}^2 \right)
  - 2 \partial_j \partial_k
 \right] T_k &=&
 2 \epsilon_{jkl}~ \partial_0 \partial_k T_{0l}~ .
 \end{eqnarray*}
All their solutions satisfy at the same time the
D'Alembert equation, therefore the standard decomposition
in plain waves can be used with subsequently extracting
the positive- and negative-frequency components
\cite{BSh}.

In momentum representation the basis of the solutions is
described as follows. Since the hamiltonian does not mix
longitudinal and transverse field configurations, they
stay independent.

There are similar {\em longitudinal} massless excitations
of scalar and pseudoscalar type, with $T_{0j}(p)$ or
$T_j(p)$ parallel to $p_j$;~ $p_0$=$ \sqrt{ \vec{p}^2}$.
The corresponding secondarily quantized field can be
written through the creation and annihilation operators
as
\begin{eqnarray*} T_{0j}(x) = \int
 \FR {{\rm d}^3\vec{p}~p_j} {2(2\pi p_0)^{3/2}}
 \Bigl\{ -~ a^\dagger(-\vec{p})~ \exp
  \left[ {\rm i}
   \left( p_0 x_0 - \vec{p}\cdot\vec{x} \right)
  \right]
 \Bigr. \\
 \Bigl. +~ a(\vec{p})~ \exp
  \left[ -{\rm i}
   \left( p_0 x_0 + \vec{p}\cdot\vec{x} \right)
  \right]
 \Bigr\}, \end{eqnarray*}
which reduces the hamiltonian (\ref{H}) to $\int {\rm
d}^3\vec{p}~ p_0~ a^\dagger(\vec{p})~ a(\vec{p})$ after
the normal ordering.

The two {\em transverse} waves involve both vector and
axial-vector fields of equal magnitude, being orthogonal
to $\vec{p}$ and to each other. The fields are
represented as follows
\begin{eqnarray*} T_{0j}(x) = \int
 \FR {{\rm d}^3\vec{p}} {2(2\pi)^{3/2}\sqrt{p_0}}
 \Bigl\{ b^\dagger(-\vec{p})~ n_j(-\vec{p})~ \exp
  \left[ {\rm i}
   \left( p_0 x_0 - \vec{p}\cdot\vec{x} \right)
  \right]
 \Bigr. \\
 \Bigl. +~ b(\vec{p})~ n_j(\vec{p})~ \exp
  \left[ -{\rm i}
   \left( p_0 x_0 + \vec{p}\cdot\vec{x} \right)
  \right]
 \Bigr\}, \\
T_j(x) = \int \FR {{\rm d}^3\vec{p}~\epsilon_{jkl}~p_k}
 {2(2\pi p_0)^{3/2}}
 \Bigl\{ -~ b^\dagger(-\vec{p})~ n_l(-\vec{p})~ \exp
  \left[ {\rm i}
   \left( p_0 x_0 - \vec{p}\cdot\vec{x} \right)
  \right]
 \Bigr. \\
 \Bigl. +~ b(\vec{p})~ n_l(\vec{p})~ \exp
  \left[ -{\rm i}
   \left( p_0 x_0 + \vec{p}\cdot\vec{x} \right)
  \right]
 \Bigr\}, \end{eqnarray*}
where $\vec{n}(\vec{p})\cdot\vec{p}=0$. However, the
hamiltonian on these solutions of the equations of motion
turns into zero, that is the model possesses unusual
zero-energy excitations.

It can easily be seen that the square form in the
lagrangian (\ref{L}) for our tensor fields is
non-degenerate (as opposed to the gauge fields
\cite{Notof}). The causal propagator is well defined
\begin{eqnarray*}
&& \left< T_{\mu\nu}(-p)~ T_{\alpha\beta}(p) \right> =
 \FR {\rm i} {p^2 + {\rm i} 0}
 \Pi_{\mu\nu\,\alpha\beta}(p), \\
&& \Pi_{\mu\nu\,\alpha\beta}(p) = \fr 1 2
 \left( g_{\mu\alpha} g_{\nu\beta}
  - g_{\mu\beta} g_{\nu\alpha}
 \right) \\*
&& ~~~~-\FR{ g_{\mu\alpha} p_\nu p_\beta
 +g_{\nu\beta} p_\mu p_\alpha
 -g_{\mu\beta} p_\nu p_\alpha
 -g_{\nu\alpha} p_\mu p_\beta } {p^2}, \\
&& \Pi_{\mu\nu\,\rho\sigma}(p)~
 \Pi_{\rho\sigma\,\alpha\beta}(p) = \fr 1 2
 \left( g_{\mu\alpha} g_{\nu\beta}
  - g_{\mu\beta} g_{\nu\alpha}
 \right). \end{eqnarray*}
Its tensorial structure just repeats the structure of the
kinetic operator in the lagrangian.

This completes the quantization of the free antisymmetric
tensor matter fields. It is worth mentioning that the
contribution of the quartic potential term of
eq.(\ref{L}) to the hamiltonian can be rewritten as (with
the matrix notation in Lorentz indices used for brevity)
$$ - \fr 1 4 q
 \left( \fr 1 2 \tr^2 T^2 - 2 \tr~ T^4 \right)
 = q
 \left[ \fr 1 2 \left( T_{0j}^2 - T_j^2 \right)^2
  + 2 \left( T_{0j} T_j \right)^2
 \right]. $$
It is evidently nonnegative-definite if $q$$\ge$0. Thus,
the existence of the vacuum should not be violated by the
self-interaction.

\section{One-loop renormalizations} \indent

By the power-counting rules the lagrangian (\ref{L})
seems to be renormalizable. However, it is well known
that in an axial-vector gauge theory there exists the
Adler -- Bell--Jackiw anomaly which leads to
non-conservation of the axial vector current and can
destroy the gauge invariance of counterterms and
renormalizability since the three-loop level. To avoid
this difficulty, we have to adjust the set of the fields
that contribute to the anomaly at one loop so that their
contributions canceled. The simplest way to achieve this
is to introduce a partner for every charged particle with
the opposite axial-gauge charge.

Thus, the cancellation of the anomaly will be guaranteed
if we add another spinor $\chi$ with the charge $-$$h$
instead of $h$ and, as a partner for $T_{\mu\nu}$,
another antisymmetric tensor field $U_{\mu\nu}$ with the
charge $-$2$h$. Identities (\ref{id}) generalize to two
fields as
$$ \til{T}_{\mu\lambda} U_{\lambda\nu}
 + \til{U}_{\mu\lambda} T_{\lambda\nu} =
 \fr 1 2 g_{\mu\nu} \tr~ (T \til{U}). $$

The doubling of the fields gives rise to the appearance
of new possible interactions. Along with the kinetic and
the minimal gauge terms, which look the same as in
eq.(\ref{L}) with $h$$\to$$-$$h$, the most general
gauge-invariant terms allowed to be added are
\begin{eqnarray} &{\cal L}_{\rm add.} =
 z~ \overline{\chi} \sigma_{\mu\nu} U_{\mu\nu} \chi
 + \fr 1 4 r \left( \fr 1 2 \tr^2 U^2 - 2\tr~U^4 \right)
 + \fr 1 2 s \tr^2 (T U)& \nonumber\\*
&+ v
 \left[ \fr 1 4 \tr~T^2 \tr~U^2
  - \tr \left( T^2 U^2 \right)
 \right] + w
 \left[ \fr 1 4 \tr~T^2 \tr~U^2
  - \tr \left( T U T U \right)
 \right],~~~& \label{L2} \end{eqnarray}
besides the mixed mass term $m^2 \tr~(T U)$ which we do
not want to introduce since it generates tachion states,
violating the positivity of the free hamiltonian. It will
not be generated in perturbation theory if we do not
introduce it from the very beginning. On the other hand,
in the minimal subtraction renormalization scheme, which
we use, renormalizations of the dimensionless couplings
do not depend on any masses. The mass term may be
essential when we shall consider the spontaneous symmetry
breaking in a more elaborate extended model.

Without slightly complicating the calculations of Feynman
diagrams, we can provide both the spinors $\psi$ and
$\chi$ with an isotopic index runnung through $n$ values.
This trivial generalization does not break the
cancellation of the anomaly.

In our one-loop calculations we use the standard Feynman
gauge. The Feynman rules for eqs.~(\ref{L}) and
(\ref{L2}) are presented in the appendix.

To unambiguously evaluate the divergent contributions in
one loop, we can apply the regularization by dimensional
reduction \cite{RDR}.

Here are our results for the renormalization-group
$\beta$ functions and anomalous dimensions of the fields,
obtained with the aid of the computer program {\em FORM}
for analytic evaluation:
\begin{eqnarray} 16~ \pi^2 \gamma_{_T} &=&
 \fr 4 3 n y^2 + \fr 4 3 h^2, ~~~~~~~
 16~ \pi^2 \gamma_{_U} ~=~ \fr 4 3 n z^2 + \fr 4 3 h^2,
 \label{gT} \\
16~ \pi^2 \gamma_\psi &=& h^2 - 6 y^2, ~~~~~~~~~~~~~~
 16~ \pi^2 \gamma_\chi ~=~ h^2 - 6 z^2, \label{gpsi} \\
\beta_{h^2} &=& \gamma_{_A} h^2 ~=~ (16~ \pi^2)^{-1}
 \left( \fr 8 3 n - 6 \right) h^4, \label{bh2} \\
16~ \pi^2 \beta_{y^2} &=&
 \left[ \fr{10}3 h^2 + \left( \fr 4 3 n - 12 \right) y^2
 \right] y^2, \label{by2} \\
16~ \pi^2 \beta_{z^2} &=&
 \left[ \fr{10}3 h^2 + \left( \fr 4 3 n - 12 \right) z^2
 \right] z^2, \label{bz2} \\
16~ \pi^2 \beta_q &=& \fr{13}9 q^2 + \fr 8 3 n y^2 q
 + \fr{14}3 h^2 q + 32~ n y^4 + 256~ h^4 \nonumber\\*
&& +~ v^2 + 2 v w + 5 w^2 - v s + w s + \fr 1 2 s^2,
 \label{bq} \\
16~ \pi^2 \beta_r &=& \fr{13}9 r^2 + \fr 8 3 n z^2 r
 + \fr{14}3 h^2 r + 32~ n z^4 + 256~ h^4 \nonumber\\*
&& +~ v^2 + 2 v w + 5 w^2 - v s + w s + \fr 1 2 s^2,
 \label{br} \\
16~ \pi^2 \beta_v &=& \fr{10}3 v^2 + \fr 8 3 v w
 + \fr 2 3 w^2 + \fr 2 3 v s + \fr 1 6 s^2 \nonumber\\*
&& + \fr 1 9 (2 v + 2 w - s) \left( q + r \right)
 + \fr 4 3 n v \left( y^2 + z^2 \right) \nonumber\\*
&& +~ 4 w h^2 - \fr{10}3 v h^2 - 2 s h^2 + \fr{256}3 h^4,
 \label{bv} \\
16~ \pi^2 \beta_w &=&
 \fr 1 6 (4 w + s) \left( q + r \right)
 + \fr 4 3 n w \left( y^2 + z^2 \right)
 + \fr 2 3 w h^2 + 2 s h^2,~~~~~~ \label{bw}\\
16~ \pi^2 \beta_s &=& -~ 8 v w + 8 v s - 8 w^2 + 12~ w s
 - 4 s^2 + \fr 1 3 (4 w + s) \left( q + r \right)
 \nonumber\\*
&& + \fr 4 3 n s \left( y^2 + z^2 \right) + 16~ w h^2
 - \fr{10}3 s h^2. \label{bs} \end{eqnarray}
In the Feynman gauge the ultraviolet divergencies in the
Yukawa vertices happen to cancel, which leaves only the
propagator contributions in eqs.~(\ref{by2}) and
(\ref{bz2}).

The most interesting fact is that at $n$=1 and 2 the
gauge charge exhibits asymptotically free behavior,
eq.(\ref{bh2}), which is due to the negative contribution
of the antisymmetric tensor matter fields (that of the
spinors is positive as in the usual quantum
electrodynamics, while the abelian axial-vector field
possesses no gauge self-interaction).

As $n$$<$3.5 or $n$$>$9, there exists a special
renormalization-group solution, consistent with
eqs.~(\ref{bh2}) -- (\ref{bz2}), which makes the Yukawa
charges proportional to the gauge charge
$$ y^2 = z^2 = \left( 2 - \FR {11} {9-n} \right) h^2. $$
However, the explicit test of eqs.~(\ref{bq}) --
(\ref{bs}) with the aid of the computer program {\em
Mathematica} shows that at $n$=1, 2, and 3 there are no
consistent special solutions with proportional real
values for all the rest of the couplings. Thus, the
self-interaction of the tensor fields does not
asymptotically diminish in the ultraviolet limit (as it
does not diminish for ordinary scalar matter fields in
the $\varphi^4$ theory). This can make the asymptotic
freedom of the gauge charge in our toy model unstable
with respect to higher-order corrections, when at the
third loop a contribution of the quartic self-couplings
to $\beta_{h^2}$ will appear. There still remains a hope,
however, that supplementing other fields may generate
complete one-charge special solutions, respecting
asymptotic freedom to all orders of perturbation theory
and possibly related to a higher symmetry or
supersymmetry.

Thus, we have demonstrated the possibility of introducing
a new type of matter fields into the gauge quantum field
theory. Some of their properties, like zero-energy free
waves and asymptotic freedom in an abelian model, are
quite unusual. Ahead stay further investigations into
non-abelian models, spontaneous symmetry breaking, and
the physics associated with extending the standard
electroweak theory to include antisymmetric tensor matter
fields, which would substantially modify the Higgs
sector.

{\bf Acknowledgement.} L.V.Avdeev is grateful to the
Physics Department of the University of Pisa, where this
work was finished, for kind hospitality, and to the
I.N.F.N. for financial support.

\renewcommand\thesubsection{Appendix.}
\subsection{Feynman rules for antisymmetric tensor
 matter fields}

\setlength\unitlength{.8mm}

\begin{eqnarray*}
&\begin{picture}(28,7)(-14,-1) 
 \put(-6,-.5){\line(1,0){12}}
 \put(-6,.5){\line(1,0){12}}
 \put(-4,3){$\leftarrow$$p$}
 \put(-15,-2){$T_{\mu\nu}$}
 \put(8,-2){$T_{\alpha\beta}$}
\end{picture}& \Rightarrow~
\begin{picture}(28,7)(-14,-1) 
 \put(-6,-.5){\line(1,0){12}}
 \put(-6,.5){\line(1,0){12}}
 \put(-5,0){\circle*{.5}}
 \put(-3,0){\circle*{.5}}
 \put(-1,0){\circle*{.5}}
 \put(1,0){\circle*{.5}}
 \put(3,0){\circle*{.5}}
 \put(5,0){\circle*{.5}}
 \put(-4,3){$\leftarrow$$p$}
 \put(-15,-2){$U_{\mu\nu}$}
 \put(8,-2){$U_{\alpha\beta}$}
\end{picture} ~~\Rightarrow
\FR{\rm i} {p^2+{\rm i}0} \Pi_{\mu\nu\,\alpha\beta}(p),\\
&\begin{picture}(26,7)(-13,-1) 
 \put(-5,0){\oval(2,2)[b]}
 \put(-3,0){\oval(2,2)[t]}
 \put(-1,0){\oval(2,2)[b]}
 \put(1,0){\oval(2,2)[t]}
 \put(3,0){\oval(2,2)[b]}
 \put(5,0){\oval(2,2)[t]}
 \put(-4,3.5){$\leftarrow$$p$}
 \put(-14,-2){$A_\mu$}
 \put(7,-2){$A_\nu$}
\end{picture}& \Rightarrow~
- \FR {\rm i} {p^2+{\rm i}0} g_{\mu\nu}~,\\
&\begin{picture}(22,7)(-11,-1) 
 \put(-6,0){\line(1,0){12}}
 \put(-1,0){\vector(-1,0)0}
 \put(-1,3){$p$}
 \put(-11,-1){$\psi$}
 \put(8,-1){$\overline{\psi}$}
\end{picture}& \Rightarrow~
\begin{picture}(22,7)(-11,-1) 
 \put(-6,0){\line(1,0)3}
 \put(-2,0){\line(1,0)4}
 \put(3,0){\line(1,0)3}
 \put(-1,0){\vector(-1,0)0}
 \put(-1,3){$p$}
 \put(-11,-1){$\chi$}
 \put(8,-1){$\overline{\chi}$}
\end{picture} ~~\Rightarrow
 \FR {{\rm i}~ p_\mu \gamma_\mu} {p^2+{\rm i}0},\\
&\begin{picture}(28,12)(-14,-2) 
 \put(-6,-.5){\line(1,0){12}}
 \put(-6,.5){\line(1,0){12}}
 \put(0,0){\circle*2}
 \put(0,1.5){\oval(2,2)[l]}
 \put(0,3.5){\oval(2,2)[r]}
 \put(0,5.5){\oval(2,2)[l]}
 \put(0,7.5){\oval(2,2)[r]}
 \put(2,5){$A_\lambda$}
 \put(-15,-2){$T_{\mu\nu}$}
 \put(8,-2){$T_{\alpha\beta}$}
 \put(-6,-4){$\leftarrow$}
 \put(-5,-7.5){$p$}
 \put(1,-4){$\rightarrow$}
 \put(2,-7.5){$q$}
\end{picture}& \Rightarrow
 \fr 1 2 h ~ ( p_\mu \epsilon_{\lambda\nu\alpha\beta}
 - p_\nu \epsilon_{\lambda\mu\alpha\beta}
 - g_{\lambda\alpha} p_\rho \epsilon_{\rho\beta\mu\nu}
 + g_{\lambda\beta} p_\rho \epsilon_{\rho\alpha\mu\nu}\\*
&& ~~~~~~~ +~ q_\alpha \epsilon_{\lambda\beta\mu\nu}
 - q_\beta \epsilon_{\lambda\alpha\mu\nu}
 - g_{\lambda\mu} q_\rho \epsilon_{\rho\nu\alpha\beta}
 + g_{\lambda\nu} q_\rho \epsilon_{\rho\mu\alpha\beta}
 ),\\
&\begin{picture}(28,12)(-14,-2) 
 \put(-6,-.5){\line(1,0){12}}
 \put(-6,.5){\line(1,0){12}}
 \put(-4.5,0){\circle*{.5}}
 \put(-2.5,0){\circle*{.5}}
 \put(2.5,0){\circle*{.5}}
 \put(4.5,0){\circle*{.5}}
 \put(0,0){\circle*2}
 \put(0,1.5){\oval(2,2)[l]}
 \put(0,3.5){\oval(2,2)[r]}
 \put(0,5.5){\oval(2,2)[l]}
 \put(0,7.5){\oval(2,2)[r]}
 \put(2,5){$A_\lambda$}
 \put(-15,-2){$U_{\mu\nu}$}
 \put(8,-2){$U_{\alpha\beta}$}
 \put(-6,-4){$\leftarrow$}
 \put(-5,-7.5){$p$}
 \put(1,-4){$\rightarrow$}
 \put(2,-7.5){$q$}
\end{picture}& \Rightarrow
 - \fr 1 2 h ~ ( p_\mu \epsilon_{\lambda\nu\alpha\beta}
 - p_\nu \epsilon_{\lambda\mu\alpha\beta}
 - g_{\lambda\alpha} p_\rho \epsilon_{\rho\beta\mu\nu}
 + g_{\lambda\beta} p_\rho \epsilon_{\rho\alpha\mu\nu}\\*
&& ~~~~~~~~ +~ q_\alpha \epsilon_{\lambda\beta\mu\nu}
 - q_\beta \epsilon_{\lambda\alpha\mu\nu}
 - g_{\lambda\mu} q_\rho \epsilon_{\rho\nu\alpha\beta}
 + g_{\lambda\nu} q_\rho \epsilon_{\rho\mu\alpha\beta}
 ),\\
&\begin{picture}(20,12)(-10,-1) 
 \put(-6,0){\line(1,0){12}}
 \put(-5,0){\vector(-1,0)0}
 \put(3,0){\vector(-1,0)0}
 \put(0,0){\circle*2}
 \put(0,1.5){\oval(2,2)[l]}
 \put(0,3.5){\oval(2,2)[r]}
 \put(0,5.5){\oval(2,2)[l]}
 \put(0,7.5){\oval(2,2)[r]}
 \put(2,5){$A_\mu$}
 \put(-11,-1){$\overline{\psi}$}
 \put(8,-1){$\psi$}
\end{picture}& \Rightarrow~
 {\rm i}~ h~ \gamma_5~ \gamma_\mu~, ~~~~~~
\begin{picture}(20,12)(-10,-1) 
 \put(-6,0){\line(1,0)3}
 \put(-2,0){\line(1,0)4}
 \put(3,0){\line(1,0)3}
 \put(-5.5,0){\vector(-1,0)0}
 \put(3.5,0){\vector(-1,0)0}
 \put(0,0){\circle*2}
 \put(0,1.5){\oval(2,2)[l]}
 \put(0,3.5){\oval(2,2)[r]}
 \put(0,5.5){\oval(2,2)[l]}
 \put(0,7.5){\oval(2,2)[r]}
 \put(2,5){$A_\mu$}
 \put(-11,-1){$\overline{\chi}$}
 \put(8,-1){$\chi$}
\end{picture} ~\Rightarrow
 - {\rm i}~ h~ \gamma_5~ \gamma_\mu~, \\
&\begin{picture}(28,10)(-14,-2) 
 \put(-4,6){\oval(4,4)[lb]}
 \put(-4,2){\oval(4,4)[rt]}
 \put(0,2){\oval(4,4)[b]}
 \put(4,2){\oval(4,4)[lt]}
 \put(4,6){\oval(4,4)[rb]}
 \put(-.7,0){\line(-1,-1)5}
 \put(0,-.7){\line(-1,-1)5}
 \put(0,-.7){\line(1,-1)5}
 \put(.7,0){\line(1,-1)5}
 \put(0,0){\circle*2}
 \put(-13,2){$A_\lambda$}
 \put(-15,-5){$T_{\mu\nu}$}
 \put(7,2){$A_\rho$}
 \put(7,-5){$T_{\alpha\beta}$}
\end{picture}& \Rightarrow~
\begin{picture}(28,10)(-14,-2) 
 \put(-4,6){\oval(4,4)[lb]}
 \put(-4,2){\oval(4,4)[rt]}
 \put(0,2){\oval(4,4)[b]}
 \put(4,2){\oval(4,4)[lt]}
 \put(4,6){\oval(4,4)[rb]}
 \put(-.7,0){\line(-1,-1)5}
 \put(0,-.7){\line(-1,-1)5}
 \put(-1.5,-1.5){\circle*{.5}}
 \put(-3,-3){\circle*{.5}}
 \put(-4.5,-4.5){\circle*{.5}}
 \put(0,-.7){\line(1,-1)5}
 \put(.7,0){\line(1,-1)5}
 \put(1.5,-1.5){\circle*{.5}}
 \put(3,-3){\circle*{.5}}
 \put(4.5,-4.5){\circle*{.5}}
 \put(0,0){\circle*2}
 \put(-13,2){$A_\lambda$}
 \put(-15,-5){$U_{\mu\nu}$}
 \put(7,2){$A_\rho$}
 \put(7,-5){$U_{\alpha\beta}$}
\end{picture} ~\Rightarrow~ 4~ {\rm i}~ h^2 \Bigl[\Bigr.
 (g_{\mu\alpha} g_{\nu\beta} - g_{\mu\beta} g_{\nu\alpha}
 ) g_{\lambda\rho} \\*[2mm]
&& ~~~~~~~
 -~ g_{\mu\alpha} (g_{\nu\lambda} g_{\beta\rho}
  + g_{\nu\rho} g_{\beta\lambda})
 + g_{\mu\beta} (g_{\nu\lambda} g_{\alpha\rho}
  + g_{\nu\rho} g_{\alpha\lambda}) \\*
&& ~~~~~~
 +~ g_{\nu\alpha} (g_{\mu\lambda} g_{\beta\rho}
  + g_{\mu\rho} g_{\beta\lambda})
 - g_{\nu\beta} (g_{\mu\lambda} g_{\alpha\rho}
  + g_{\mu\rho} g_{\alpha\lambda}) \Bigl.\Bigr], \\
&\begin{picture}(20,12)(-10,-1) 
 \put(-6,0){\line(1,0){12}}
 \put(-5,0){\vector(-1,0)0}
 \put(3,0){\vector(-1,0)0}
 \put(0,0){\circle*2}
 \put(-.5,0){\line(0,1)8}
 \put(.5,0){\line(0,1)8}
 \put(2,5){$T_{\mu\nu}$}
 \put(-11,-1){$\overline{\psi}$}
 \put(8,-1){$\psi$}
\end{picture}& \Rightarrow~
 {\rm i}~ y~ \sigma_{\mu\nu}~,~~~~~~
\begin{picture}(20,12)(-10,-1) 
 \put(-6,0){\line(1,0)3}
 \put(-2,0){\line(1,0)4}
 \put(3,0){\line(1,0)3}
 \put(-5.5,0){\vector(-1,0)0}
 \put(3.5,0){\vector(-1,0)0}
 \put(0,0){\circle*2}
 \put(-.5,0){\line(0,1)8}
 \put(.5,0){\line(0,1)8}
 \put(0,2.5){\circle*{.5}}
 \put(0,4.5){\circle*{.5}}
 \put(0,6.5){\circle*{.5}}
 \put(2,5){$U_{\mu\nu}$}
 \put(-11,-1){$\overline{\chi}$}
 \put(8,-1){$\chi$}
\end{picture} ~\Rightarrow~
 {\rm i}~ z~ \sigma_{\mu\nu}~, \\
&\begin{picture}(30,10)(-15,-2) 
 \put(-5.7,-5){\line(1,1){10.7}}
 \put(-5,-5.7){\line(1,1){10.7}}
 \put(5,-5.7){\line(-1,1){10.7}}
 \put(5.7,-5){\line(-1,1){10.7}}
 \put(0,0){\circle*2}
 \put(-18,2){$T_{\mu_1\nu_1}$}
 \put(-18,-5){$T_{\mu_2\nu_2}$}
 \put(7,2){$T_{\mu_4\nu_4}$}
 \put(7,-5){$T_{\mu_3\nu_3}$}
\end{picture}& \Rightarrow~ {\rm i}~ q~ \prod_{j=1}^4
\left[ \fr 1 2
 \left( g_{\mu_j\alpha_j} g_{\nu_j\beta_j}
  - g_{\mu_j\beta_j} g_{\nu_j\alpha_j}
 \right)
\right] \times \\*
&& \Bigl[\Bigr.
\fr 1 3 ( g_{\alpha_1\alpha_2} g_{\beta_1\beta_2}
  g_{\alpha_3\alpha_4} g_{\beta_3\beta_4}
 + g_{\alpha_1\alpha_3} g_{\beta_1\beta_3}
  g_{\alpha_2\alpha_4} g_{\beta_2\beta_4} \\*
&&~ +~ g_{\alpha_1\alpha_4} g_{\beta_1\beta_4}
  g_{\alpha_2\alpha_3} g_{\beta_2\beta_3} )~
 - \fr 4 3 ( g_{\beta_1\alpha_2} g_{\beta_2\alpha_3}
  g_{\beta_3\alpha_4} g_{\beta_4\alpha_1} \\*
&&~~ +~ g_{\beta_1\alpha_3} g_{\beta_3\alpha_2}
  g_{\beta_2\alpha_4} g_{\beta_4\alpha_1}
 + g_{\beta_1\alpha_2} g_{\beta_2\alpha_4}
  g_{\beta_4\alpha_3} g_{\beta_3\alpha_1} ) \Bigl.\Bigr],
\\
&\begin{picture}(30,10)(-15,-2) 
 \put(-5.7,-5){\line(1,1){10.7}}
 \put(-5,-5.7){\line(1,1){10.7}}
 \put(-4.5,-4.5){\circle*{.5}}
 \put(-3,-3){\circle*{.5}}
 \put(-1.5,-1.5){\circle*{.5}}
 \put(1.5,1.5){\circle*{.5}}
 \put(3,3){\circle*{.5}}
 \put(4.5,4.5){\circle*{.5}}
 \put(5,-5.7){\line(-1,1){10.7}}
 \put(5.7,-5){\line(-1,1){10.7}}
 \put(4.5,-4.5){\circle*{.5}}
 \put(3,-3){\circle*{.5}}
 \put(1.5,-1.5){\circle*{.5}}
 \put(-1.5,1.5){\circle*{.5}}
 \put(-3,3){\circle*{.5}}
 \put(-4.5,4.5){\circle*{.5}}
 \put(0,0){\circle*2}
 \put(-18,2){$U_{\mu_1\nu_1}$}
 \put(-18,-5){$U_{\mu_2\nu_2}$}
 \put(7,2){$U_{\mu_4\nu_4}$}
 \put(7,-5){$U_{\mu_3\nu_3}$}
\end{picture}& \Rightarrow~ {\rm i}~ r~ \prod_{j=1}^4
\left[ \fr 1 2
 \left( g_{\mu_j\alpha_j} g_{\nu_j\beta_j}
  - g_{\mu_j\beta_j} g_{\nu_j\alpha_j}
 \right)
\right] \times \\*
&& \Bigl[\Bigr.
\fr 1 3 ( g_{\alpha_1\alpha_2} g_{\beta_1\beta_2}
  g_{\alpha_3\alpha_4} g_{\beta_3\beta_4}
 + g_{\alpha_1\alpha_3} g_{\beta_1\beta_3}
  g_{\alpha_2\alpha_4} g_{\beta_2\beta_4} \\*
&&~ +~ g_{\alpha_1\alpha_4} g_{\beta_1\beta_4}
  g_{\alpha_2\alpha_3} g_{\beta_2\beta_3} )~
 - \fr 4 3 ( g_{\beta_1\alpha_2} g_{\beta_2\alpha_3}
  g_{\beta_3\alpha_4} g_{\beta_4\alpha_1} \\*
&&~~ +~ g_{\beta_1\alpha_3} g_{\beta_3\alpha_2}
  g_{\beta_2\alpha_4} g_{\beta_4\alpha_1}
 + g_{\beta_1\alpha_2} g_{\beta_2\alpha_4}
  g_{\beta_4\alpha_3} g_{\beta_3\alpha_1} ) \Bigl.\Bigr],
\\
&\begin{picture}(30,10)(-15,-2) 
 \put(-5.7,-5){\line(1,1){10.7}}
 \put(-5,-5.7){\line(1,1){10.7}}
 \put(5,-5.7){\line(-1,1){10.7}}
 \put(5.7,-5){\line(-1,1){10.7}}
 \put(4.5,-4.5){\circle*{.5}}
 \put(3,-3){\circle*{.5}}
 \put(1.5,-1.5){\circle*{.5}}
 \put(1.5,1.5){\circle*{.5}}
 \put(3,3){\circle*{.5}}
 \put(4.5,4.5){\circle*{.5}}
 \put(0,0){\circle*2}
 \put(-18,2){$T_{\mu_1\nu_1}$}
 \put(-18,-5){$T_{\mu_2\nu_2}$}
 \put(7,2){$U_{\mu_4\nu_4}$}
 \put(7,-5){$U_{\mu_3\nu_3}$}
\end{picture}& \Rightarrow~ {\rm i}~ \prod_{j=1}^4
\left[ \fr 1 2
 \left( g_{\mu_j\alpha_j} g_{\nu_j\beta_j}
  - g_{\mu_j\beta_j} g_{\nu_j\alpha_j}
 \right)
\right] \times \\*
&& \Bigl[\Bigr.
\fr 1 2 s~ ( g_{\alpha_1\alpha_3} g_{\beta_1\beta_3}
  g_{\alpha_2\alpha_4} g_{\beta_2\beta_4}
 + g_{\alpha_2\alpha_3} g_{\beta_2\beta_3}
  g_{\alpha_1\alpha_4} g_{\beta_1\beta_4} ) \\*
&&~ +~ v~ ( g_{\alpha_1\alpha_2} g_{\beta_1\beta_2}
  g_{\alpha_3\alpha_4} g_{\beta_3\beta_4}
 - 2 g_{\beta_1\alpha_2} g_{\beta_2\alpha_3}
  g_{\beta_3\alpha_4} g_{\beta_4\alpha_1} \\*
&& -~ 2 g_{\beta_2\alpha_1} g_{\beta_1\alpha_3}
  g_{\beta_3\alpha_4} g_{\beta_4\alpha_2} )
 ~ +~ w~ ( g_{\alpha_1\alpha_2} g_{\beta_1\beta_2}
  g_{\alpha_3\alpha_4} g_{\beta_3\beta_4} \\*
&&~~ -~ 2 g_{\beta_1\alpha_3} g_{\beta_3\alpha_2}
  g_{\beta_2\alpha_4} g_{\beta_4\alpha_1}
 - 2 g_{\beta_2\alpha_3} g_{\beta_3\alpha_1}
  g_{\beta_1\alpha_4} g_{\beta_4\alpha_2} ) \Bigl.\Bigr].
\end{eqnarray*}

\pagebreak[4]

\end{document}